\documentclass[aps,prl,notitlepage,reprint]{revtex4-1}
\usepackage[utf8]{inputenc}
\usepackage{amsmath}
\usepackage{amsfonts}
\usepackage{amssymb}
\usepackage{graphicx}
\usepackage{color}
\usepackage{soul}

\newcommand{\Pe}{\mathrm{Pe}}
\newcommand{\cri}{_\text{cr}}
\newcommand{\Dr}{D_\text{r}}
\newcommand{\vhi}{\varphi}
\newcommand{\mean}[1]{\langle #1\rangle}
\newcommand{\gam}{\gamma}

\begin{document}

\title{Critical behavior of active Brownian particles}
\author{Jonathan Tammo Siebert, Florian Dittrich, Friederike Schmid, Kurt Binder, Thomas Speck, and Peter Virnau}
\affiliation{Johannes Gutenberg University Mainz, Department of Physics, 55128 Mainz}
\date{\today}
\begin{abstract}
  We study active Brownian particles as a paradigm for genuine non-equilibrium phase transitions. Access to the critical point in computer simulations is obstructed by the fact that the density is conserved. We propose a modification of sampling finite-size fluctuations and successfully test this method for the 2D Ising model. Using this model allows us to determine accurately the critical point of two dimensional active Brownian particles at $\Pe\cri=40(2)$, $\phi\cri=0.597(3)$. Based on this estimate, we study the corresponding critical exponents $\beta$, $\gamma/\nu$, and $\nu$. Our results are incompatible with the 2D-Ising exponents, thus raising the question whether there exists a corresponding non-equilibrium universality class.
\end{abstract}

\maketitle

The notion ``active matter'' encompasses a wide range of systems and phenomena at the border of physics, chemistry, and biology that share a common trait: they are out of thermal equilibrium due to \emph{local} dissipation stemming from the \emph{directed motion} of its constituents. Examples range from actomyosin~\cite{Koenderink:2009,Schaller:2010,Sanchez:2012} (actin filaments driven by molecular motors) to swimming bacteria~\cite{Zhang:2010,Wensink:2012a} to colloidal particles propelled by a multitude of mechanisms~\cite{Paxton:2006,Hong:2007,Palacci:2010,Jiang:2010,Buttinoni:2013,Dreyfus:2005,Peddireddy:2012,Herminghaus:2014,Wang:2012}. The interplay of interactions with this persistent motion leads to a variety of collective dynamic behaviors such as swarming~\cite{Wensink:2012}, turbulent motion~\cite{Dombrowski:2004}, giant number fluctuations~\cite{Ramaswamy:2003,Narayan:2007}, and clustering~\cite{Bialke:2012,Das:2014,Trefz:2016,Siebert:2017}. In the language of statistical physics, this behavior often can be characterized as "phases" with abrupt changes depending on external parameters such as temperature and density.

Such phase transitions have been investigated intensively, and much has been learned from the study of minimal model systems. The arguably simplest model that shows an order-disorder phase transition ending in a critical point is the Ising model of a lattice of spins interacting with their nearest neighbors. In particular the understanding of critical points has sparked intensive research that has cumulated in the development of tools such as the renormalization group and finite-size scaling in computer simulations that have found application in a wide range of problems. Specifically the system-size dependence of order parameter fluctuations and the crossing of cumulants has proven to be very successful~\cite{Luijten:1999,Luijten:2002,Campostrini:2001,Watanabe:2012}. Denoting the order parameter as $m$, the ratio 
\begin{equation}
  \label{eq:Q}
  Q_\ell = \mean{m^2}^2/\mean{m^4}
\end{equation}
becomes independent of system size $\ell$ exactly at the critical point. Hence, plotting this quantity for several values of $\ell$ allows to locate the critical point from the intersection with high accuracy.\par
Universal behavior and scaling invariance are not restricted to passive systems but are also observed in systems driven away from thermal equilibrium. A well-studied paradigm constitutes the KPZ equation, originally proposed for the evolution of interfaces~\cite{Kardar:1986}. Regarding phase transitions, previous studies have focused on non-equilibrium effects on the critical point and the critical exponents of an underlying \emph{equilibrium} phase transition. Examples include two dimensional and three dimensional Ising models under shear~\cite{Winter:2010,Hucht:2009,Angst:2012,Hucht:2012} and active versions of the Ising model~\cite{Solon:2015a,Attanasi:2014a}. In the context of active particles, the influence of self-propulsion on the gas-liquid transition in the continuous Asakura-Oosawa model~\cite{Trefz:2017} (with alignment interactions) and in a Lennard-Jones fluid~\cite{Prymidis:2016} (without alignment interactions) have been determined.

While driving a system featuring a passive transition influences its critical behavior, genuine non-equilibrium transitions without a passive counterpart are much less studied. Active Brownian particles (ABPs) have emerged as a minimal model showing such a transition: the coexistence of dilute and dense regions in the absence of cohesive forces~\cite{Bialke:2015a,Fily:2012,Redner:2013,Solon:2015,Stenhammar:2014,Wysocki:2014,Siebert:2017}. While the binodal lines away from the critical point have already been determined~\cite{Bialke:2015a,Siebert:2017} with good accuracy, the precise position of the critical point remains unknown. The close resemblance with passive phase separation suggests that density fluctuations of ABPs become scale invariant in the vicinity of the critical point. In the following, we employ extensive computer simulations to shed first light on the critical behavior of ABPs and the intriguing possibility of a novel non-equilibrium universality class.


To be specific, we simulate $N$ particles moving in $d=2$ dimensions in a box with periodic boundaries. The coupled equations of motion read
\begin{equation}
  \label{eq:dyn}
  \dot{\mathbf r}_k = -\nabla_kU + \frac{\Pe}{d_{\mathrm{BH}}/\sigma}\left(
    \begin{array}{c}
      \cos\vhi_k \\ \sin\vhi_k
    \end{array}\right) + \sqrt{2}\mathbf R_k
\end{equation}
with normal distributed Gaussian noise $\mathbf R_k$ and potential energy $U$ modeling short-range repulsion  with effective hard disk diameter $d_\text{BH}$ and Lennard Jones length $\sigma$. Every particle has an orientation, the evolution of which is described by the angle $\vhi_k$ undergoing free rotational diffusion with diffusion coefficient $\Dr$. Particles are propelled along this orientation with constant speed. Throughout, we employ dimensionless quantities with the speed given by the Pecl\'{e}t number $\Pe$. Further details can be found in the Supplementary Information~\cite{sm}.

\begin{figure}[t]
  \includegraphics{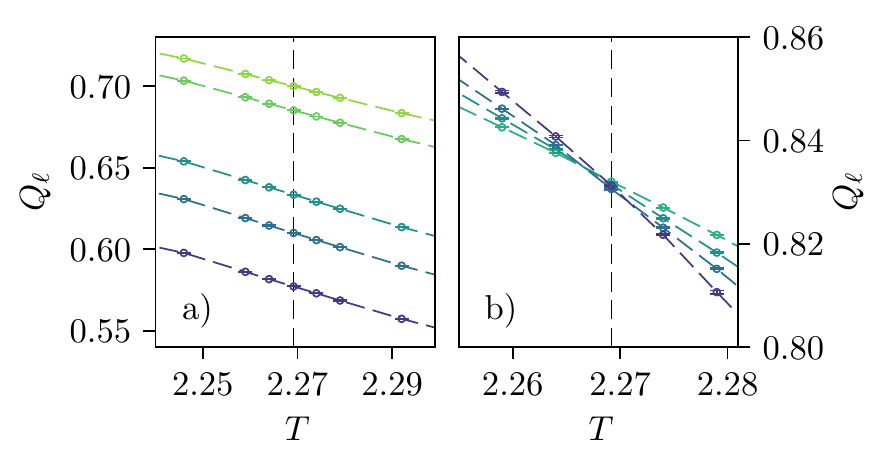}\\
  \includegraphics{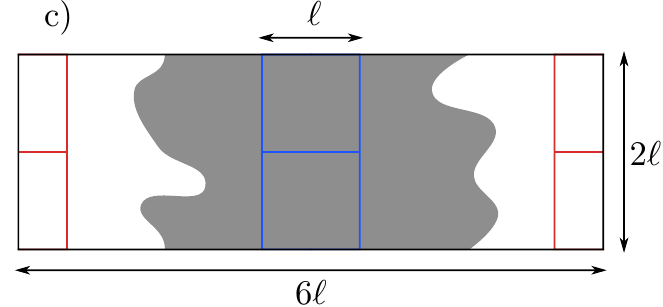}\\
  \caption{Critical temperature of the 2D Ising model. a)~Cumulant ratio $Q_\ell(T)$ as a function of temperature $T$ for different subbox sizes $\ell=5,6,10,12,15$ (top to bottom) do not cross applying the original block-magnetization distribution method on an underlying canonical simulation. As discussed in Ref.~\cite{Rovere:1993}, the cumulants do not intersect due to the presence of interfaces. b)~Using the modified block-magnetization method, $Q_\ell(T)$-curves for different $\ell=8,10,12,15$ (same color scale as in a)) now cross very close to the critical temperature $T_\text{c}\approx2.269185$ (indicated by the dashed vertical line) even if the underlying simulation is canonical. c)~Schematic representation of the simulation box used for the cumulant analysis. Simulations are done at medium packing fractions in an elongated box with an edge length ratio of 1:3. This results in a slab geometry where the slab is always aligned with the short axis. Two subboxes are then placed at the center of mass. The other two subboxes are shifted by $3\ell$ in $x$ direction.}
  \label{fig:isingQL}
\end{figure}

To determine critical points in passive fluids and suspensions, best practice is to conduct numerical simulations in the grand canonical ensemble with the total number of particles fluctuating~\cite{Bruce:1992,Wilding:1995,Wilding:1997}. In driven systems, this option is not (yet) available due to the lack of a rigorous free energy. An alternative strategy to sample density fluctuations are block-density-distribution methods~\cite{Binder:1981,Binder:1987,Rovere:1988,Rovere:1990,Rovere:1993}. By subdividing a simulation box into smaller subboxes, we allow every subbox to have a fluctuating particle number while the remaining system effectively acts as a particle reservoir. Especially in three dimensional off-lattice systems this approach has proven to be very successful~\cite{Watanabe:2012}. Even though it only works for a rather small range of intermediate subbox sizes it provides accurate results in equilibrium~\cite{Watanabe:2012} as well as non-equilibrium systems~\cite{Trefz:2017}. Nonetheless, there are severe draw-backs especially in two dimensions. For off-lattice systems, \emph{e.g.} a Lennard-Jones fluid in two dimensions, the method seems to work to some extent, but for the 2D Ising model, it completely fails if the underlying simulation takes place in the canonical ensemble~\cite{Rovere:1993}. This failure is demonstrated in Fig.~\ref{fig:isingQL}a). The plot shows the cumulant ratio $Q_\ell$~[Eq. (\ref{eq:Q})] for the subbox magnetization $m$ as a function of temperature $T$ averaged over independent runs for different subbox lengths $\ell$ using the original block-magnetization-distribution method~\cite{Binder:1981,Luijten:1999,Luijten:2002}. The curves do not cross over a large temperature range around the critical temperature.

This failure can be traced to the biased measurement of the order parameter distribution in the subboxes, which does not reproduce the grand canonical distribution due to the over-expression of interfaces~\cite{Rovere:1993}. To solve this problem, we propose an improved block-distribution method~\cite{sm}. In a nutshell, we exploit the stability of interfaces in a finite system (even in the vicinity of a critical point) to sample subboxes away from the interface. By simulating an elongated box with aspect ratio $1:3$, we force the system into a slab geometry, see Fig.~\ref{fig:isingQL}c). Although, close to the critical point, fluctuations increase such that bubbles or even rifts can appear. Going further into the homogeneous region, the slab eventually dissolves. Four subboxes of size $\ell\times\ell$ are placed in the box, two at the center of mass and two shifted by $3\ell$ in $x$ direction. Aside from avoiding the interfaces, the necessary simulations of systems with different sizes then allow to also eliminate the second length scale that is introduced by the size of the surrounding simulation box in the original method. Including only the indicated boxes into the calculation of the magnetization, the new method is indeed able to predict the critical point of the 2D Ising model with remarkable accuracy. Below the critical point, \emph{i.e.}, in the phase-separated region, the $Q_\ell(T)$-curves are ordered going from large values for large subboxes to small values for small subboxes. At the critical temperature the curves now cross and at even higher temperatures, \emph{i.e.}, in the homogeneous region, they invert their order. This shows that our new method indeed allows to circumvent the main problems of the original block-magnetization-distribution method.

\begin{figure}[!t]
  \includegraphics{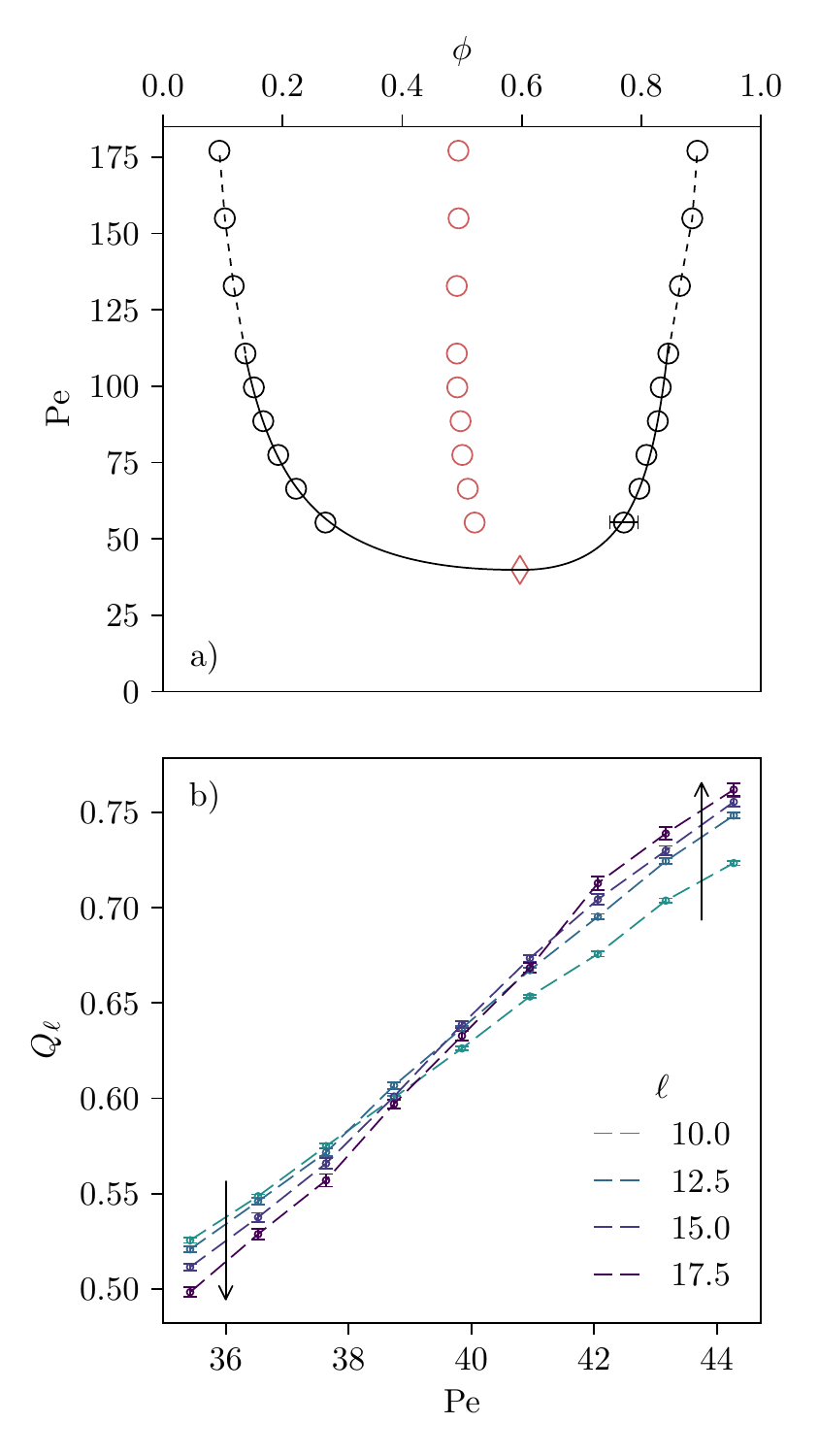}
  \caption{Critical point of active Brownian particles. a)~Coexisting packing fractions $\phi$~\cite{Bialke:2015a,Siebert:2017} showing our estimate for the critical point as a red diamond. The rectilinear diameter is shown as red circles. For all points without errorbars, the statistical uncertainty is smaller than the symbol size. Points far from the critical point ($\Pe>120$) are connected by a dashed line as a guide for the eye. For points close to critical point, the gas and liquid branch are both fitted with a power law, where the exponent $\beta=0.45$ is the best estimate arising from our analysis of the critical exponents. b)~Cumulant intersection analysis for ABPs. A crossing of $Q_\ell(\Pe)$ [Eq.~\eqref{eq:Q}] for all system sizes $\ell$ can be seen between $\Pe\simeq38$ and $\Pe\simeq42$, giving an estimate of the critical point of $\Pe\cri=40(2)$. Error bars are estimated from independent runs. The dashed lines are again only included as guides to the eye.}
  \label{fig:newCrossing}
\end{figure}

Encouraged by these results, we now return to the active Brownian particles. Analogously to the Ising system, we study simulation boxes with aspect ratio of $1:3$ and then evaluate subboxes in the center of the dense and the dilute slab (cf. Fig.~\ref{fig:isingQL}c). In place of the magnetization we employ the subbox density fluctuations $m=\rho_\ell-\mean{\rho_\ell}$ away from the average density and vary the propulsion speed $\Pe$. Here $\rho_\ell=N_\ell/\ell^2$ with $N_\ell$ the fluctuating number of particles in a subbox with edge length $\ell$. The resulting curves for $Q_\ell(\Pe)$ with values of $\ell$ between $10$ and $17.5$ are shown in Fig.~\ref{fig:newCrossing}b). To make contact to the physics of hard spheres and previous estimates of the phase diagram~\cite{Stenhammar:2014,Bialke:2015a,Siebert:2017}, we use the packing fraction $\phi=\rho\pi d_\text{BH}^2/4$ instead of the density. Similarly to the Ising system, the curves show the correct ordering above ($\Pe\ge42.1$) and below ($\Pe\le37.6$) a putative critical point. Between $\Pe=37.6$ and $\Pe=42.1$ the curves cross. This is already a very remarkable result as it supports that scaling laws as known from equilibrium finite size scaling are valid also in this non-equilibrium system. Outside this interval, the points corresponding to different edge lengths are clearly separated, whereas within this intermediate interval the points' uncertainties do not allow to distinguish between them, which in turn indicates that $\Pe\cri$ lies within this interval. Note that even after eliminating the additional scaling variable of box length over subbox length there are still successive intersections over this region. Also, even though every point corresponds to between 58-174 independent runs that are used to determine the average of $Q_\ell(\Pe)$, the resulting uncertainties in $Q_\ell$ are still notable. Nonetheless, this analysis allows to estimate the critical point to be at $\Pe\cri=40(2)$. To estimate the critical density, we average the mean packing fractions $\mean{\phi_\ell}$ over all subbox sizes, all independent runs, and over all Peclet numbers between $37.6$ and $42.1$. This results in an estimate of $\phi\cri=0.597(3)$. As this is an average over different subbox sizes and Peclet numbers, the uncertainty is given as the standard deviation of the density for all subbox sizes and Peclet numbers each averaged over all respective runs.

Even though our new method gives a much more accurate and reliable result, we also checked its consistency with the original block-density-distribution method, which, regardless of its shortcomings, still gives an estimate of the critical point~\cite{sm}. This estimate is compatible with the results of the modified method excluding interfaces (albeit of course less precise). Furthermore, it is possible to determine a lower bound for the critical speed based on the divergence of the static structure factor, as well as an upper bound by analyzing the cluster size distribution. Both bounds agree well with our current estimate. Altogether, we can conclude that the critical point in ABPs is located at $\Pe\cri = 40(2)$ and $\phi\cri = 0.597(3)$. This point is shown as a red diamond in the phase diagram in Fig.~\ref{fig:newCrossing}a). Interestingly, the critical point in ABPs does not lie on the linear extension of the rectilinear diameter, which is shown as red circles. While this is rather uncommon in equilibrium systems, a similar behavior has been found for other non-equilibrium transitions as well~\cite{Prymidis:2016,Trefz:2017}.

\begin{figure*}[t]
  \includegraphics{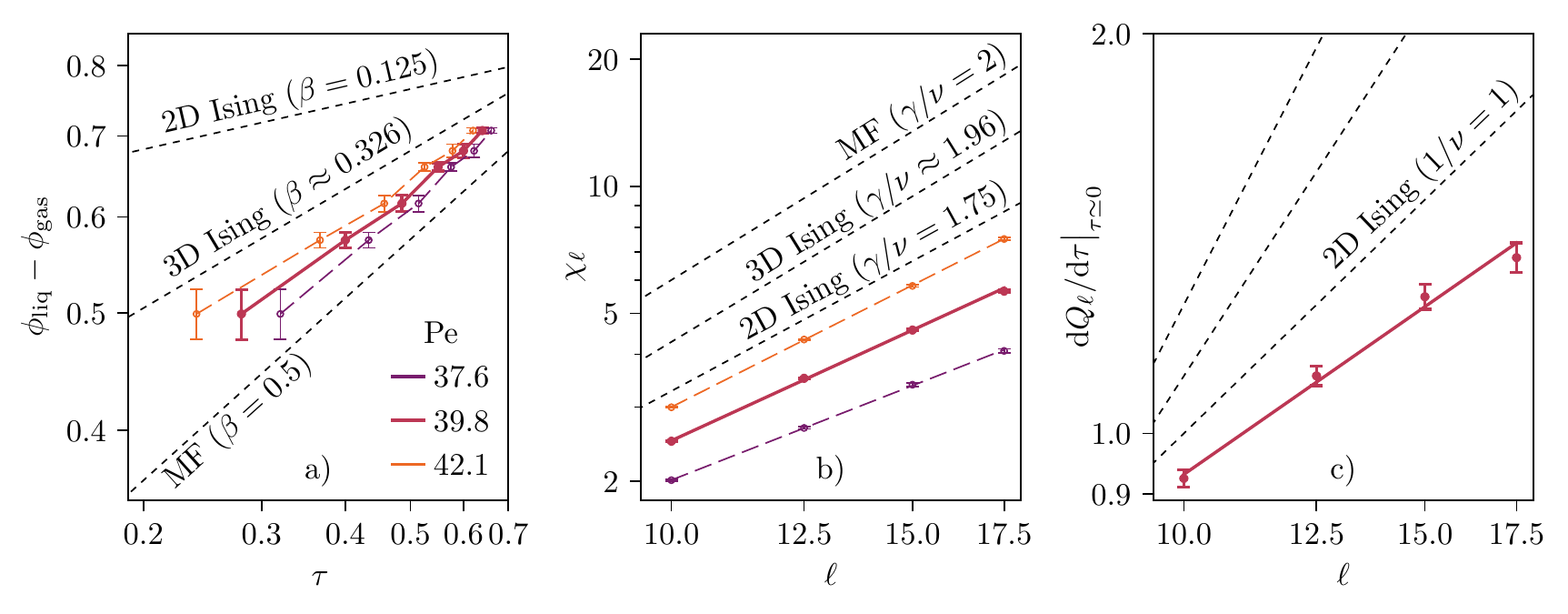}
  \caption{Determination of critical exponents (thick lines correspond to the estimate of $\Pe\cri$, the dashed lines show lower (purple) and upper (orange) bounds). a)~Log-log-plot of the order parameter $\mean{m}\propto\phi_{\mathrm{liq}}-\phi_{\mathrm{gas}}$ \emph{vs.} the distance to the critical point $\tau$ [Eq.~\eqref{eq:tau}] for different estimates of the critical speed. Connecting lines are shown as guides to the eye. The order parameter exponent $\beta$ is given by the slope of the curve. Ising and MF slopes are shown as dashed lines for reference. b)~Log-log-plot of the particle number fluctuations' dependence on the subbox size. Their slope corresponds to $\gam/\nu$ with the susceptibility exponent $\gamma$ and the correlation length exponent $\nu$. Again, Ising and MF slopes are shown as dashed lines for reference. c)~Log-log-plot of $Q_{\ell}$'s slope against the subbox size. The derivative $\mathrm{d}Q_{\ell}/\mathrm{d}\tau\big|_{\tau\simeq0}$ is determined by fitting a linear function to $Q_{\ell}(\tau)$ in the critical region. The slope of the curve in the log-log-plot corresponds to $1/\nu$~\cite{sm}. 2D-Ising ($\nu=1$), 3D-Ising ($\nu\simeq0.63$), and MF ($\nu=1/2$) universality are shown for reference as dashed lines in increasing order of steepness.}
  \label{fig:exponents}
\end{figure*}


We now extract numerical estimates for the critical exponents allowing insight into the critical behavior and the universality class of ABPs. For this purpose, we define the dimensionless distance
\begin{equation}
  \label{eq:tau}
  \tau = \frac{\Pe^{-1}-\Pe\cri^{-1}}{\Pe\cri^{-1}}
\end{equation}
to the critical point, generalizing the usual expression by treating the propulsion speed as an inverse temperature. First, we turn to the order parameter exponent $\beta$. In the phase separated region, but still close to the critical point, one expects a power-law increase of the mean order parameter $\mean{m}\propto\tau^\beta$, see Fig.~\ref{fig:exponents}a). To account for the uncertainty in the determination of $\Pe\cri$, we show three curves corresponding to our best estimate of $\Pe\cri\simeq40$ as well as (generous) lower and upper bounds of $37.6$ and $42.1$, respectively. All three curves show a reasonably linear behavior within the uncertainties of the order parameter. Instead of a fit, as guides we show the slopes corresponding to 2D Ising ($\beta=1/8$), 3D Ising ($\beta\approx0.326$), and MF ($\beta=1/2$). For all reasonable estimates of $\Pe\cri$, the slope of the resulting curve is significantly lower than that of the 2D Ising universality behavior.

Both the finite-size behavior of $Q_\ell$ and the behavior of the order parameter approaching the critical point indicate that usual scaling arguments are applicable, with scale-free density fluctuations at the critical point. This implies the existence of a correlation length $\xi$ that diverges as $\xi\sim\tau^{-\nu}$ with exponent $\nu$. Assuming that the susceptibility
\begin{equation}
  \label{eq:chi}
  \chi_{\ell} = \frac{\mean{(N_\ell-\mean{N_\ell})^2}}{\mean{N_\ell}}
\end{equation}
diverges as $\chi_{\infty}\sim\tau^{-\gam}$ in the infinite size limit, one derives the relation $\chi_{\ell}=\chi_0(\ell/\xi)\xi^{\gam/\nu}=\ell^{\gam/\nu} \tilde\chi(x)$ with scaling function $\tilde\chi(x)$ replacing $\tau$ by $\xi$ and assuming a prefactor $\chi_0$ that depends on system size only through the ratio $x=\ell/\xi$. Since $\xi$ is bound by the dimension of the full box $2\ell$ and the quotient of that length and the subbox height is fixed, close to the critical point the scaling function saturates to a constant value and we can extract the ratio of $\gamma/\nu$ from the slope of the calculated $\chi_{\ell}$ plotted against $\ell$ in Fig.~\ref{fig:exponents}b). We see that the resulting slope is again smaller than that expected for 2D Ising ($\gamma/\nu=1.75$), and even farther from 3D Ising ($\gamma/\nu\approx1.96$) or MF ($\gamma/\nu=2$).\par
Finally, we turn to the dependence of $Q_{\ell}$'s derivative with respect to the distance from the critical point around criticality: $\mathrm{d}Q_{\ell}/\mathrm{d}\tau\big|_{\tau\simeq0}$. It is expected to have a power law dependence on the system size $\ell$ with exponent $1/\nu$~\cite{sm}. To estimate the derivative, we fit $Q_{\ell}(\tau)$ in the region were we estimated the critical point [$\Pe=40(2)$] with a linear function. Its dependence on the system size is shown in Fig.~\ref{fig:exponents}c). The slope of this curve is significantly lower than that expected for 2D Ising ($1/\nu=1$), indicating that $\nu>1$. Using rough estimates for the critical exponents ($\beta\simeq0.45$, $\nu\simeq1.4$, and $\gamma\simeq2.1$) a tentative check of the scaling relation $\gamma+2\beta=2\nu$ shows that it is approximately satisfied, which, given the numerical uncertainties, serves as a provisional check of consistency.


Within a mean-field treatment of the equations~\eqref{eq:dyn}, the qualitative phase behavior of ABPs is indeed recovered with a critical point characterized by the expected mean-field exponents~\cite{Speck:2015}. The relevant non-linearity $\sim\rho^4$ is that of the Ising class for short-range interactions. Hence, in the presence of additive noise one would expect that the critical point also falls into the Ising universality class. Somewhat surprisingly (given the strong resemblance with ordinary phase separation), our numerical results indicate that this might not be the case. Extracting critical exponents from numerical data crucially depends on the precise determination of the critical point. An exact determination, which would allow a definite answer to the question of the existence of an ``active matter'' universality class, is still precluded by the statistical uncertainties in the determination of the critical point. Nevertheless, our best estimate $\Pe\cri\simeq 40(2)$ for the critical speed implies that all exponents do not agree with the corresponding 2D Ising values, cf. Fig.~\ref{fig:exponents}.

To conclude, employing a novel method we were able to determine the critical point of ABPs to be at $\Pe\cri=40(2)$ and $\phi\cri=0.597(3)$. Moreover, we have provided numerical evidence that the universality class might not agree with Ising 2D universality despite the strong qualitative agreement with passive liquid-gas phase separation. This is somewhat unexpected and we hope that these results will stimulate further research into the theoretical underpinning of scale invariance in active matter and genuine non-equilibrium phase transitions.


\begin{acknowledgments}
  JTS, TS, and PV gratefully acknowledge financial support by the DFG within priority program SPP 1726 (Grants No. SP1382/3-2 and VI 237/5-2). ZDV Mainz is acknowledged for computing time on the MOGON supercomputers.
\end{acknowledgments}


\bibliography{references}

\end{document}


\title{{\small Supplementary Information} \\ Critical behavior of active Brownian particles}
\author{Jonathan Tammo Siebert, Florian Dittrich, Friederike Schmid, Kurt Binder, Thomas Speck, and Peter Virnau}
\affiliation{Johannes Gutenberg University Mainz, Department of Physics, 55128 Mainz}
\date{\today}

\maketitle

\section{Model}

The dynamics of active Brownian Particles are governed by the coupled equations of motion
\begin{equation}
  \dot{\mathbf r}_k = -\nabla_kU + \frac{\Pe}{d_{\mathrm{BH}}/\sigma}\left(
    \begin{array}{c}
      \cos\vhi_k \\ \sin\vhi_k
    \end{array}\right) + \sqrt{2}\mathbf R_k
\end{equation}
and
\begin{equation}
  \dot{\vhi}_k = \sqrt{2 D_{\mathrm{r}}} T_k
\end{equation}
with Peclet number $\Pe=(3v_0)/(d_{\mathrm{BH}}D_{\mathrm{r}})$ and independent and normal distributed Gaussian noises $\mathbf R_k$ and $T_k$. The interactions
\begin{equation}
  U(\{\mathbf{r}_i\}) = \sum_{i<j}u_\text{WCA}(|\mathbf{r}_i-\mathbf{r}_j|)
\end{equation}
are modeled via the strongly repulsive Weeks-Chandler-Anderson pair potential~\cite{Weeks:1971}
\begin{equation}
  u_{\text{WCA}}(r) = \begin{cases}4\epsilon \left(r^{-12}-r^{-6}+\frac{1}{4}\right) & r < 2^{1/6} \\
    0 & r \ge 2^{1/6}.
  \end{cases}
  \label{eq:potential}
\end{equation}
As units of length and time we choose $\sigma$ and $\sigma^2/D$ with bare diffusion coefficient $D$, respectively. The interaction strength $\epsilon=100$ is chosen to only allow for small overlaps. The resulting effective diameter~\cite{Barker:1967} is $d_{\mathrm{BH}}\approx 1.10688$. The rotational diffusion constant is given by $D_{\mathrm{r}}=3/d_{\mathrm{BH}}^2$. The equations are then integrated by an Euler scheme except for the data on the larger boxes that were used to determine the static structure factor, which were integrated using a higher order predictor-corrector scheme~\cite{Kloeden:1999}.

\section{Improved block-distribution method}

As the overexpression of interfaces seems to be the main problem in the original scheme, we try to include only subboxes that do not contain an interface. By simulating an elongated box with side length ratio $1:3$, we force the system into a slab geometry, exploiting the stability of interfaces in finite systems. We then place four subboxes with side lengths $\ell$ in the system. To ensure a proper sampling of both phases but also the exclusion of the interface region we place two subboxes aligned parallel to the slab in the middle of each, the dense and the dilute region. The dense boxes are positioned around the center of mass in $x$ direction~\cite{Bai:2008}. The dilute boxes are then shifted by $3\ell$.

As a proof of concept, we test this method using the example of the two dimensional Ising model. In contrast to the original approach for which the curves corresponding to different system sizes do not cross at all (Fig.~1 a) of the manuscript), the new method is indeed able to predict the critical point with remarkable accuracy. Fig.~1 b) of the manuscript shows $Q_{\ell}(T)$ with subbox magnetization $m$ for subbox sizes of $\ell = 8, 10, 12, 15$. Below the critical point, i.e. in the phase separated region, the $Q_{\ell}(T)$-curves are ordered going from large values for large subboxes to small values for small subboxes. At the critical temperature (which is indicated by the dashed vertical line) the curves cross and at even higher temperatures, i.e., in the homogeneous region, they invert their order. This shows that our new method indeed allows to accurately predict the critical temperature in the two dimensional Ising model, circumventing the main problems of the original block-magnetization-distribution method. Excluding the interface region, the measured distribution is much closer to the true grand canonical distribution. Furthermore, simulating different sized systems for the different subbox sizes allows to eliminate the additional scaling variable of box length over subbox length occurring in the original block-distribution method.

\begin{figure}[t]
	\includegraphics{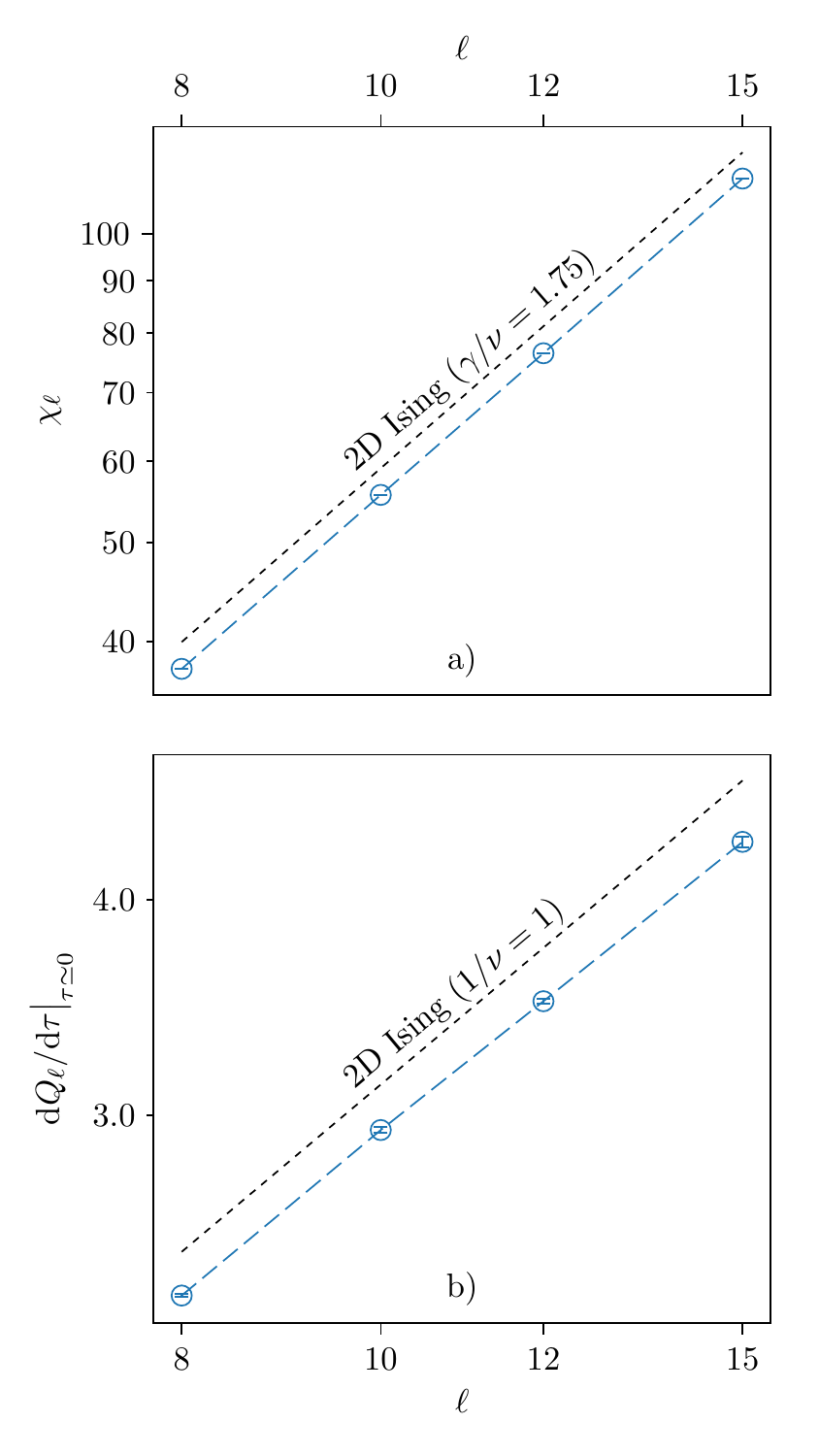}
	\caption{a)~Log-log-plot of the susceptibility \emph{vs.} the subbox size. The slope corresponds to $\gam/\nu$ (cf. Fig.~3b) of the manuscript). The dashed line shows corresponds to the literature value, fitting perfectly to the data's slope. b)~Log-log-plot of $Q_{\ell}$'s slope against the subbox size. The derivative $\mathrm{d}Q_{\ell}/\mathrm{d}\tau\big|_{\tau\simeq0}$ is determined by fitting a linear function to $Q_{\ell}(\tau)$ in the critical region. The slope of the curve in the log-log-plot corresponds to $1/\nu$~\cite{sm} (cf. Fig.~3c) of the manuscript). 2D-Ising ($\nu=1$) universality is shown for reference as dashed lines, fitting to the data's slope within its uncertainties.}
	\label{fig:exponentsIsing}
\end{figure}

In addition to the determination of the critical point, data from the analysis can be used to estimate critical exponents (cf. Fig.~3) of the manuscript). Fig.~\ref{fig:exponentsIsing} demonstrates that the analysis is working nicely in the case of the 2D Ising model, for which the slopes in the log-log-plot reproduce the exponents well within the uncertainties of the data.

\section{Qualitative justification for crossing of cumulants}

A crucial aspect of this subsystem method is that in the regime of the ordered phase two of the four $\ell \times \ell$ subboxes are centered in the middle of the ``liquid'' domain, which has the linear dimension $(2\ell) \times (3\ell)$ (Fig. 1 of the manuscript). This constraint must be maintained not only at low temperatures but also in the critical region, noting that (in the grandcanonical ensemble) the order parameter distribution in $d=2$ dimensions still has a pronounced double-peak structure, throughout the critical region, and the two peaks merge in a single peak above $T_c$ when the correlation length $\xi$ has become much smaller than the linear dimension $2\ell$ of the system. For this reason, it is useful to choose the coordinate system such that the origin coincides with the center of mass of the system (the Ising model is then considered as lattice gas, particles have mass unity). The geometry then looks as shown in Figure 1 a) of the manuscript.\par
For systems in the Ising universality class, critical correlations grow isotropically. Thus, the growth of the correlation length $\xi$ is limited by the smaller linear dimension $2\ell$ in the chosen geometry. The average distance of the liquid-vapor interfaces from the boundaries of the measurement boxes is $\ell$, and when $\xi$ is distinctly smaller than $2\ell$, the typical length scale $w$ of interfacial fluctuations in the $d=2$ Ising model is, from capillary wave theory and exact solutions \cite{Abraham:1986,Binder:1995}\par
\begin{equation} \label{eq1}
w \propto \sqrt{2\ell\xi}
\end{equation}
where the prefactor is of the order unity. Eq.~(\ref{eq1}) implies that for $\xi \ll 2\ell$ the ``measurements'' of the density in the subboxes cannot be affected by interfacial fluctuations, and it is reasonable to assume that the fluctuations in the liquid subboxes are independent from the fluctuations in the vapor subboxes. Of course, the fluctuations in the two liquid subboxes are not independent of each other, because they interact across their boundaries in $y$ directions twice (each subbox has the same upper and lower neighbor subbox, because of the periodic boundary conditions in $y$-direction).\par
However, in $x$-direction there are intervening $2\ell \times 2\ell$ regions on both sides of the subboxes (half filled by liquid regions, half by vapor) which act as ``particle reservoirs'' for the particle number fluctuations in the ``measurement boxes''. This situation is not identical to the grandcanonical ensemble of statistical mechanics, but qualitatively similar (in the grandcanonical ensemble, one considers a subbox which can exchange particles with an infinitely large reservoir at the same average density \cite{Landau:1958a}). Thus, when $2\ell \gg \xi$, one can in principle divide each $\ell \times \ell$ subbox into many weakly interacting subsystems to conclude that the distribution of density in the subbox must be (approximately) Gaussian, due to the law of large number \cite{Landau:1958a,Binder:1981,Rovere:1993,Roman:1997a,Roman:1998,Sengupta:2000}\par
\begin{equation} \label{eq2}
p^{\rm liqiud}_{\rm subbox} (\rho) \propto \exp \{-(\rho-\rho^{\rm coex}_\ell)^2 \, \ell^2/(2 k_BT \chi^\ell_{\rm eff}) \}\text{,} \end{equation}
\begin{equation} \label{eq3}
p^{\rm vapor}_{\rm subbox} (\rho) \propto \exp \{-(\rho-\rho^{\rm coex}_v)^2 \, \ell^2/(2 k_BT \chi^\upsilon_{\rm eff}) \}\text{.} \end{equation}
Taking the average of all 4 subboxes yields the desired double-peak distribution, analogous to the grandcanonical ensemble.\par
For a lattice gas, we have a symmetry for the ``susceptibilities'' $\chi^\ell_{\rm eff}= \chi^\upsilon_{\rm eff}$, while no such symmetry is expected for off-lattice fluids, of course. However, it can be asserted that for the chosen geometry these effective susceptibilities typically will be smaller than their counterparts in the grandcanonical ensemble, but still of the same order of magnitude. The fact that they are smaller can be concluded from the expectation that the constraints of conserved total density in the system (we work here at an average density $\rho=\rho_{\rm crit}$, $\rho_{\rm crit}=(\rho^{\rm coex}_v+ \rho^{\rm coex}_\ell)/2=1/2$ in the lattice gas) removes some fluctuations, which still are possible in the grandcanonical ensemble. The expectation that $\chi^\ell_{\rm eff}$, $\chi^v_{\rm eff}$ are of the same order as their grandcanonical counterpart $\chi^\ell$, $\chi^v$ can be justified from the explicit computation of these quantities for $\ell \times \ell$ subsystems of $L \times L$ homogeneous systems in the canonical ensemble \cite{Roman:1997a,Roman:1998,Sengupta:2000} for various fluids. Due to the presence of two interfaces in our system, however, one cannot take over any of the results in the literature \cite{Roman:1997a,Roman:1998,Sengupta:2000} to the present case quantitatively.\par
When Eqs.~(\ref{eq2},~\ref{eq3}) hold, it follows immediately \cite{Binder:1981} that the cumulant of the density distribution (with respect to the average density $\rho_{\rm crit}$) converges to 1 for $T<T_c$ when $\ell \rightarrow \infty$, i.e.~with $\Delta \rho=\rho-\rho_{\rm crit}$ we have
\begin{equation} \label{eq5}
Q_{\ell} =\frac{\langle (\Delta \rho)^2 \rangle^2_{\ell}}{\langle (\Delta \rho)^4 \rangle_{\ell}} \quad {\underset{\ell \rightarrow \infty}{\longrightarrow}} 1 \quad .
\end{equation}
Note that this result does not hold for subboxes which contain interfaces, and hence when one fails to exclude those \cite{Rovere:1993} the finite size analysis of the density distribution no longer is straightforward.\par
Of course, when one wishes to study critical phenomena, one is not only interested in the region $T < T_c$, but one wishes to pass through the critical region well up into the region of the weakly correlated disordered phase. For $T \gg T_c$, when one still fixes the boxes which were related to the liquid (for $T \leq T_c$) at the origin, there will nevertheless be no longer any significant difference in the density distribution of any of the boxes; the $6 \ell \times 2\ell$ system  has density inhomogeneities only on scales $\xi\ll 2\ell$ at $T \gg T_c$). So averaging over all 4 subboxes again follows a Gaussian distribution, but now centered at the average density $\rho_{\rm crit}$,\par
\begin{equation} \label{eq6}
p_{\rm subbox} (\rho) \propto \exp \{-(\rho-\rho_{\rm crit})^2 \ell^2/(k_BT \chi_{\rm eff})\}
\end{equation}
and hence in this region the cumulant $Q_{\ell} \rightarrow 1/3$ as $\ell \rightarrow \infty.$
Of course, for the region near $T_c$ one can postulate (which includes additional assumptions about scaling relations at this point) the same finite size scaling hypothesis as proposed in \cite{Rovere:1993}, again referring to an average over all 4 subboxes\\
\begin{equation} \label{eq7}
p_{\rm subbox} (\rho)=\ell^{\beta/\nu} \tilde{p}\{(\rho-\rho_{\rm crit}) \ell^{\beta/\nu}, \, \ell^{1/\nu} \tau \} \quad,
\end{equation}
where $\tau=1-T/T_c$, $\beta$ and $\nu$ are the critical exponents of the Ising model, and $\tilde{p}$ is a scaling function, similar - but not identical - to the scaling function that applies in the grand-canonical ensemble.\par
Eq.~(\ref{eq7}) is essentially identical to the proposal given in \cite{Rovere:1993}, for subsystems of a canonical ensemble. However, in this paper an average over all subsystems contained in the total system was taken, not conceiving that one needs to distinguish between subsystems for the ``measurements'' and subsystems containing the interfaces, and moreover acting as particle reservoirs. Thus, although Eq.~(\ref{eq7}) was proposed earlier \cite{Rovere:1993}, this proposal really referred to a physically different situation, and also its usefulness could NOT be demonstrated previously, for systems in the canonical ensemble of statistical mechanics.\par
Note that Eq.~(\ref{eq7}) leads to the standard expressions for the moments, for instance
\begin{equation} \label{eq8}
\langle (\Delta \rho)^{2k} \rangle= \ell^{-2k \beta/\nu} f_{2k} (\ell^{1/\nu} \tau)
\end{equation}
where $f_{2k}$ is a scaling function whose explicit form is not needed, and hence
\begin{equation} \label{eq9}
Q_{\ell}=\tilde{Q}(\ell^{1/\nu} \tau) \quad,
\end{equation}
in the regime where $\ell \gg 1$ and $\xi \gg 1$, and $\ell$, $\xi$ are about of the same order ($\xi$ would become infinite for $\tau=0$, for a macroscopic system, of course). Near $\tau=0$ the scaling function $\tilde{Q}$ can be expanded as a Taylor series\\
\begin{equation} \label{eq10}
Q_{\ell} =\tilde{Q} (0) + \tilde{Q}' \ell^{1/\nu} \tau + \cdots ;
\end{equation}
the constant $\tilde{Q}(0)$ is similar (but not identical) to the corresponding constant of the grand canonical ensemble. Note that from Eqs.~(\ref{eq8}),~(\ref{eq10}) both exponents $\beta/\nu$ and $1/\nu$ can be estimated, since \par
\begin{equation} \label{eq11}
\langle (\Delta \rho)^2 \rangle \propto \ell^{-2 \beta/\nu} \quad, \quad d Q_{\ell}/d\tau\big|_{\tau=0} \, \propto \ell^{1/\nu} \quad .
\end{equation}
Eqs.~(\ref{eq11}) are identical to their counterparts in the grandcanonical ensemble, but for subsystems taken from a canonical ensemble their practical usefulness has not been shown earlier.

\section{Results of the original subsystem method}

\begin{figure}[t]
  \includegraphics{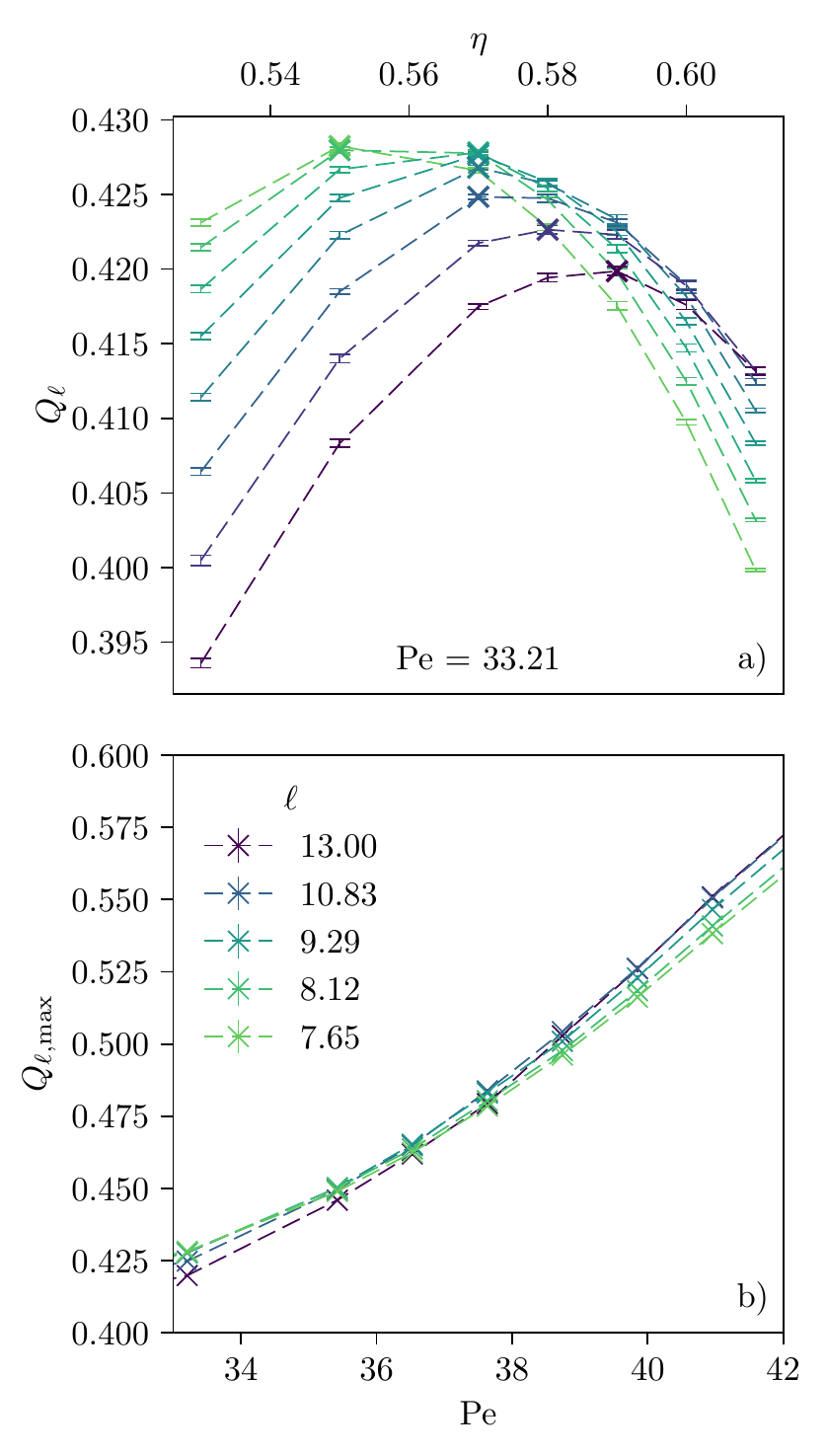}
  \caption{Original subsystem method:\quad a) Dependence of $Q_{\ell}$ in subsystems of different sizes on the overall density at a propulsion strength in the homogeneous region $\Pe=33.21$. The curves exhibit maxima at increasing densities for increasing subsystem size. As suggested in~\cite{Luijten:2002,Kim:2003,Kim:2003a} we evaluate $Q_{\ell}$ for all subsystems along their locus of maximum value (as indicated by the crosses).\quad b) Crossing of $Q_{\ell,\mathrm{max}}(\Pe)$ when going from the homogeneous to the phase separated region. Generally, the results agree with those of our modified subsystem method presented in the manuscript. Note, however, a more pronounced spread of  intersections starting at $\Pe\approx 34$ for small subsystems and going up to $\Pe\approx 41$ for the larger subsystems.}
  \label{fig:oldMethod}
\end{figure}

We also compared our estimate for the critical point of ABPs to the results of the original subsystem method \cite{Binder:1981,Binder:1987,Rovere:1988,Rovere:1990,Rovere:1993}, in which a quadratic box of side length $L$ is subdivided into $N^2$ subsystems by a regular grid of spacing $\ell=L/N$. Each of these strongly correlated subsystems is then treated as a quasi-grandcanonical system which allows to determine $Q_{\ell}$. In contrast to the modified method used in our manuscript, the medium density averaged over all subsystems is fixed. Measuring $Q_{\ell}$ at different overall densities yields maxima as seen in Figure \ref{fig:oldMethod} a) for a system in the homogeneous region. Analyzing $Q_{\ell}$ along their loci of maximum value $Q_{\ell,\mathrm{max}}(\Pe)$~\cite{Luijten:2002,Kim:2003,Kim:2003a} yields intersections similar to those of the modified method presented in the manuscript. Due to a more pronounced spread of intersections, the uncertainty is much larger though. As shown in Figure \ref{fig:oldMethod} b) these intersection move from $\Pe\approx 34$ up to $\Pe\approx 41$ starting with smaller and going to larger subsystems. Possible causes for this could be the over expression of interfaces due to the phase separation as well as the additional scaling variable $N$, which were already mentioned in the manuscript. Note that, as is common in subsystem distribution methods \cite{Binder:1981,Binder:1987,Rovere:1988,Rovere:1990,Rovere:1993,Watanabe:2012,Trefz:2017}, only a small range of subsystem sizes can be used and the selection process is somewhat empirical. The critical density can be estimated by the position of the locus of large subsystems at the critical propulsion strength. This estimate is rather rough, as the resolution of density points used is low. Despite the comparably large uncertainties and the other shortcomings of the method, its results corroborate the estimate of the critical point in the manuscript in both propulsion strength and density.

\section{Static structure factor}
\begin{figure*}[t]
	\includegraphics{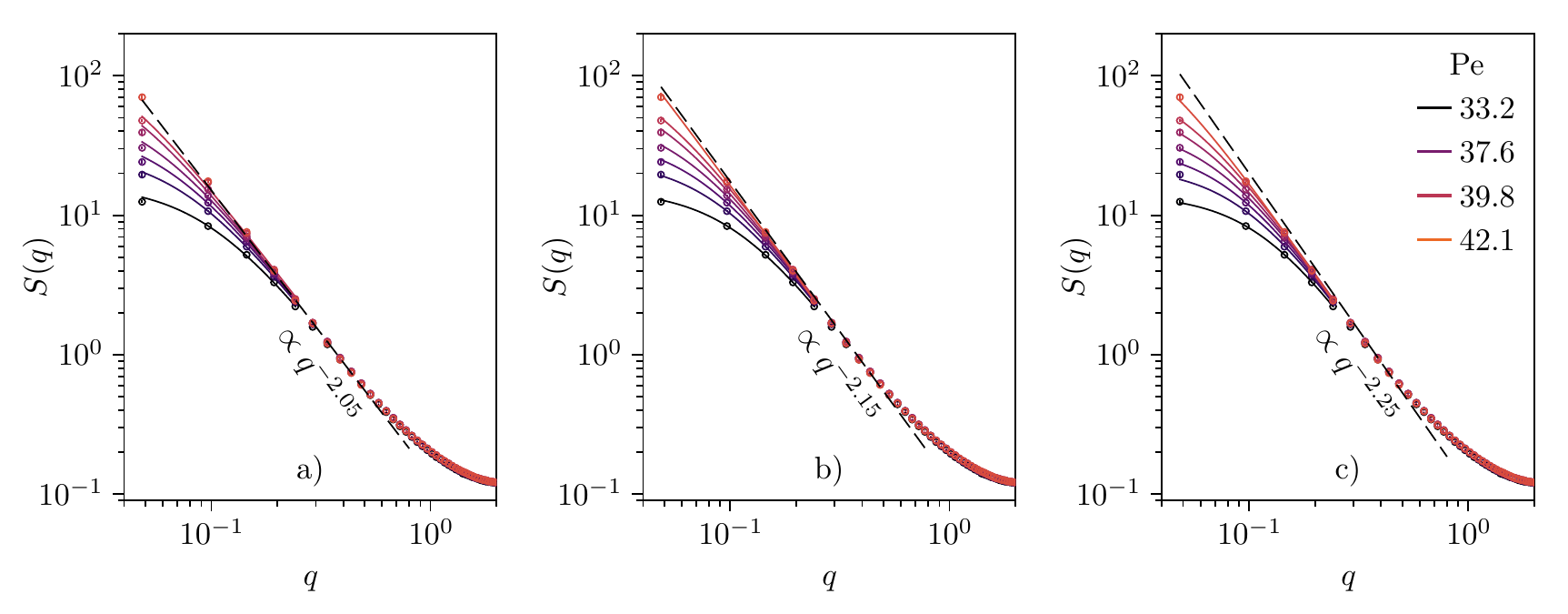}
	\caption{Static structure factor fitted for different estimates of the anomalous dimension $\eta$: a)-c) fitted with a Lorentzian as specified in Eq. (\ref{eq:lorentzian}) assuming $\eta=-0.05, -0.15, -0.25$ as different (unphysical) values for the anomalous dimension respectively. It seems that an naive extraction of correlation lengths from the low-$q$ limit of $S(q)$ determined at fixed particle number is not working as it is neglecting important finite size effects.}
	\label{fig:differentEta}
\end{figure*}

We also studied the low-$q$ limit of the static structure factor measured in a quadratic box of side length $L=130$ at a constant particle number corresponding to $\phi\simeq\phi\cri$. For sufficiently large systems it is expected to follow a Lorentzian~\cite{Fisher:1964} (see also Ref.~\cite{Fily:2012} for $\eta=0$):
\begin{equation}
	S(q)=\frac{S_0}{1+(\xi q)^{2-\eta}}\text{,}
	\label{eq:lorentzian}
\end{equation}
that can be fitted to the data to get an estimate of the correlation length $\xi$. In this case $S_0$ is a free fit parameter, $\eta$ is the (constant) anomalous dimension that determines the power law scaling $S(q)\propto q^{-2+\eta}$ for intermediary $q$ values. Note, though, that this power law is only valid for $\frac{2\pi}{\ell}\ll q \ll\frac{2\pi}{D}$, where $D$ is the typical distance between nearest neighbor particles at the chosen density. In addition one must be in the regime $\frac{2\pi}{\xi}\ll q$. Figure~\ref{fig:differentEta}a)-c) show the structure factor data, comparing it to power law slopes indicated by dashed lines. By naively fitting the low-$q$ limit of the data to Eq.~(\ref{eq:lorentzian}) assuming an anomalous dimension corresponding to the maximum slope, one finds that only $\eta<0$ is able to reproduce the measured $S(q)$ reasonably well. However, a negative anomalous dimension is not physical, as it would imply a divergence of order parameter auto-correlations with increasing distance.\par
It turns out that an naive extraction of anomalous dimensions and correlation lengths from the static structure factor in this finite system with constant particle number is not successful. It seems that the system size $L=130a$ studied in this work is still not sufficient to allow a reasonable extraction of the power law behavior $S(q)\propto q^{-2+\eta}$ for intermediary $q$ values. Instead, neglecting important finite size effects is giving unphysical results.
\bibliography{references}